\documentclass{cs21proc}

\usepackage{graphicx}
\usepackage{subfig}
\usepackage{natbib}
\usepackage{subfig}
\usepackage{float}
\usepackage{multirow,multicol}
\usepackage{footnote}
\usepackage{hyperref}
\usepackage{xcolor, colortbl}
\usepackage{lastpage}
\usepackage{verbatim}
\hypersetup{
colorlinks = true, 
urlcolor = magenta,
linkcolor = blue, 
citecolor = blue 
}
\usepackage[all]{hypcap}
\usepackage{txfonts}

\newcommand{\teff}{T_\mathrm{eff}}
\newcommand{\logg}{\mathrm{log}\,g}
\newcommand{\iraf}{\textsc{iraf}}
\usepackage{numdef}
\num\def{\lp8}{LP~877-72}
\num\def{\lp1}{LP~1033-31}
\def\arcsec{\hbox{$^{\prime\prime}$}} 
\def\lsun{$\rm{L}_{\odot}$} 
\def\msun{$\rm{M}_{\odot}$} 
\def\rsun{$\rm{R}_{\odot}$} 

\newcommand{\gsimeq}{\hbox{\raise0.5ex\hbox{$>\lower1.06ex\hbox{$\kern-0.92em{\sim}$}$}}}
\newcommand{\lsimeq}{\hbox{\raise0.5ex\hbox{$<\lower1.06ex\hbox{$\kern-0.92em{\sim}$}$}}}

\usepackage{pgffor}
\newcommand\blfootnote[1]{%
  \begingroup
  \renewcommand\thefootnote{}\footnote{#1}%
  \addtocounter{footnote}{-1}%
  \endgroup
}

\editors{A. S. Brun, J. Bouvier, P. Petit}
\publisher{Zenodo}
\conference{The 21st Cambridge Workshop on Cool Stars, Stellar Systems, and the Sun}
\conferencedate{2022}

\title{Investigation of very low mass binaries using \textit{VLT}/NaCo}
\author{Subhajeet Karmakar,$^{1,2} \dagger$
        A. S. Rajpurohit,$^{2}$
        and D. Homeier$^{3,4}$}

\affiliation{
$^{1}$ Monterey Institute for Research in Astronomy (MIRA), 200 Eighth Street, Marina, California 93933, USA\\
$^{2}$ Astronomy \& Astrophysics Division, Physical Research Laboratory, Navrangapura, Ahmedabad 380009, India\\
$^{3}$ Aperio Software Ltd., Insight House, Riverside Business Park, Stoney Common Road, Stansted, Essex, CM24 8PL, UK\\
$^{4}$ F{\"o}rderkreis Planetarium G{\"o}ttingen, 37085 G{\"o}ttingen, Germany
}

\shorttitle{Very Low Mass Binaries}
\shortauthors{Karmakar et al.}

\abs{
  \textbf{\textit{Context:}} Most stars in the galactic stellar population are low-mass stars. Very low mass (VLM) stars are a subset of the low-mass stars typically defined in terms of the stellar masses ranging from 0.6~\msun\ to the hydrogen-burning limit of about 0.075 \msun.\\ 
  \textbf{\textit{Aim:}} The observational studies of VLM binaries can provide effective diagnostics for testing the VLM formation scenarios. The small size of VLMs makes them suitable candidates to detect
planets around them in the habitable zone.\\
\textbf{\textit{Methods:}} In this work, using the high-resolution near-infrared adaptive optics imaging from the NaCo instrument installed on the Very Large Telescope, we report the discovery of a new binary companion to the M-dwarf \lp1\ and also confirm the binarity of \lp8. We have characterized both stellar systems and estimated the properties of their individual components.\\
\textbf{\textit{Results and Conclusions:}} We have found that \lp1~AB with the spectral type of M4.5+M4.5 has a projected separation of 6.7$\pm$1.3~AU. On the other hand, with the spectral type of M1+M4, the projected separation of \lp8~AB is estimated to be 45.8$\pm$0.3~AU. We further investigated the masses, surface gravity, radii, and effective temperature of the detected components. The orbital period of \lp1\ and \lp8\ systems are estimated to be $\sim$28 and $\sim$349~yr, respectively.
Our analysis suggests that there is a possibility of finding up to `two' exoplanets around LP 877-72 B. In contrast, the maximum probability of hosting exoplanets around LP 877-72 A, LP 1033-31 A, and LP 1033-31 B are estimated to be only $\sim$50\%.
}
\begin{document}

\maketitle

\blfootnote{$\dagger$ E-mails: \href{mailto:subhajeet09@gmail.com}{subhajeet09@gmail.com} (SK), \href{mailto:sk@mira.org}{sk@mira.org} (SK), \href{mailto:arvindr@prl.res.in}{arvindr@prl.res.in} (ASR), and \href{mailto:homeier@planetarium-goettingen.de}{homeier@planetarium-goettingen.de} (DH).}
\setcounter{footnote}{0}
\vspace{-1.3em}
\section{Introduction}
Low mass stars with partially or fully convective envelops show a high level of magnetic activities \citep[see][]{Pandey-15-AJ-6, Patel-16-MNRAS-6, Savanov-16-AcA-3, SavanovI-19-AstL-8, KarmakarS-19-PhDT-2, KarmakarS-21-MIRAN-1, Karmakar-16-MNRAS-8, Karmakar-17-ApJ-5, KarmakarS-22-MNRAS-4, KarmakarS-23-MNRAS}.
Recent studies have shown the magnetic activities also increase for the later spectral types \citep[e.g.,][]{Savanov-18-ARep-8, Karmakar-18-BSRSL-10, KarmakarS-19-BSRSL}.
Typically, the stars with a mass range of 0.075--0.6~\msun\ are called very low mass stars (VLMs). The VLMs are intrinsically faint due to their small size and low temperature. However, the latest studies with several ground-based surveys, including SDSS \citep[][]{YorkD-00-AJ-1} and 2MASS \citep[][]{SkrutskieM-06-AJ}, and space-based surveys including WISE \citep[][]{WrightE-10-AJ-1} have made discoveries of the VLM stars. The observational studies of the VLM stars provide effective diagnostics for testing the VLM star formation scenarios. The VLM binaries (= VLM + VLM) are of particular interest because the formation mechanisms leave their own traces on the statistical properties of the binaries, such as frequency, orbital separation, and mass-ratio distribution \citep[e.g.][]{BateM-09-MNRAS-1}. Moreover, the VLM binaries also provide the model-independent way to determine the physical properties, including masses and radii \citep[see][]{Delfosse-00-A+A-3, DupuyT-17-ApJS-4, Winters-19-AJ-4}.

In recent years the adaptive optics (AO) imaging and radial velocity (RV) survey have made a valuable contribution to the discovery of the VLM binaries. Due to their small size, VLM stars also are suitable candidates for detecting exoplanets around them in the habitable zone. 
 Recently, a few extensive exoplanet surveys such as the M2K program \citep[][]{AppsK-10-PASP},  CARMENES search for exoplanets \citep[][]{ReinersA-18-A+A-2}, and MEarth project \citep[][]{DittmannJ-16-ApJ-4} investigated exoplanets around a sizeable number of M-dwarfs.
The presence of the companion to the primary VLM star is also significant as they influence planet formation. However, due to limited detection of exoplanetary systems in the binary or multiple systems, particularly on VLM stellar systems, the study of stellar multiplicity's effect on planet formation is still statistically insignificant. In the solar neighborhood, planets in multiple-star systems are found to occur 4.5$\pm$3.2, 2.6$\pm$1.0, and 1.7$\pm$0.5 times less frequently than the single-star systems when a stellar companion is present at a distance of 10, 100, and 1000 AU, respectively. Therefore, further observations of the VLM binaries are essential.

Investigation of VLM stars with high spatial resolution imaging enabled us to discover new companions to the previously known `single' stellar system. This investigation is essential to have a better understanding of the stellar system and have a precise estimation of the binary separation for VLM stars. Significant progress has been made in this direction in the last few decades. Summarizing the results of the high spatial resolution imaging surveys of VLM stars published in dozens of articles between 1991 and 2005, \cite{Phan-Bao-05-A+A-3} found the binary frequency in the separation range 1--15 AU was about 15\%, whereas for wide binary systems (semi-major axis $>$15 AU), the binary frequency was very low, $<$1\%. Using statistical investigation utilizing a Bayesian algorithm \cite{AllenP-12-AJ} found that only 2.3\% of VLM objects have a companion in the 40--1000 AU range. While the last two sources cannot be directly compared due to different input sample sizes, the older studies might have underestimated the binary fraction or were still incomplete. Further, \cite{BaronF-15-ApJ-2} discovered 14 VLM binary systems with separations of 250--7500 AU. \cite{Galvez-Ortiz-17-MNRAS} identified 36 low-mass and VLM candidates to binary/multiple systems with separations between 200 and 92\,000 AU. These differences can be due to a continuous formation mass-dependent trend or differences in the VLM objects' formation mechanism. Further high-resolution study of VLM binaries provides valuable constraints for the formation mechanism of VLM binaries.

\section{The Objects}
In this paper, we discussed near-infrared (NIR) detection of VLM companions to two M-dwarfs \lp1\ and \lp8\ using direct imaging. Both the objects are far from the galactic plane, and they were first detected as part of HPM surveys \citep[][]{GiclasH-78-LowOB-6, LuytenW-95-yCat-1}. Since a star's proper motion (PM) is inversely proportional to its distance from the observer, a high proper motion (HPM) is a strong selection criterion for proximity. Further observations of these objects were carried out as part of several high proper-motion surveys \citep[][]{WroblewskiH-95-A+AS, PokornyR-03-A+A, SalimS-03-ApJ-3, ReyleC-04-A+A-2, SchneiderA-16-ApJ-8, KirkpatrickJ-16-ApJS-3}. The latest estimated values for proper motion of the systems are 0\arcsec $\!$.299 and 0\arcsec $\!$.305 per year for \lp1\ and \lp8, respectively \citep[][]{Muirhead.-18-AJ-3, StassunK-19-AJ-2}. Both of these objects were also studied as a part of the bright M-dwarfs \citep[][]{LepineS-11-AJ-3, FrithJ-13-MNRAS-1}, as well as the solar-neighborhoods \citep[][]{Finch-14-AJ-8, WintersJ-15-AJ-2}. We discovered \lp1\ is a VLM binary system rather than a `single' object, as known previously. Moreover, we also confirmed the binarity of \lp8\ after the \textit{Gaia} detection of the companion \citep[][]{Gaia-Collaboration-18-A+A-5}. Please find the paper \cite{KarmakarS-20-MNRAS-2} for the detailed investigation.

\section{The observatory and the AO system}
The observations were taken from the Very Large Telescope (VLT) of the European Southern Observatory (ESO, Chile). The Nasmyth Adaptive Optics System \citep[NAOS;][]{RoussetG-98-SPIE, RoussetG-00-SPIE-1} delivers diffraction-limited images to the Coud\'{e} Near Infrared Camera \citep[CONICA;][]{LenzenR-98-SPIE-1, HartungM-00-SPIE}. The combined system is popularly known as NaCo. NAOS is installed at one of the Nasmyth foci, and it corrects an f/15 beam for atmospheric turbulence and hands-on again an f/15 beam providing CONICA with diffraction-limited images. The wavefront sensing path consists of a field selector \citep[][]{SpanoudakisP-00-SPIE} and two wavefront sensors (WFS). They are located between the dichroic mirror and the WFS input focus. Please see \cite{HartungM-03-A+A-3} and references therein.

\section{Observations}
Using the VLT/NaCo instrument, two M-dwarfs \lp1\ and \lp8\ were observed under program ID 091.C-0501(B). On 2013 July 02, a total of 20 and 30 frames were observed in each of the NIR $JHK_s$ photometric bands for \lp1\ and \lp8, respectively. The exposure time for individual frames was 30~s. Each image was jittered using a jitter box of 5\arcsec\ to allow an efficient reduction of the sky background. Due to the fact that no bright adaptive optics (AO) source was available nearby, N90C10 dichroic was used, where 90\% of the light was utilized for AO and only 10\% for the science camera. S13 camera was used with a field of view of $14\arcsec \times 14\arcsec$ and a pixel scale of 13.2~mas per pixel. The airmass of \lp1\ and \lp8\ were in the range of 1.049 -- 1.051 and 1.002 -- 1.010, respectively. The seeing condition for the day of observation was also very good ($<$ 0.85\arcsec).  

\begin{figure*}[tbp]
	\centering
	\subfloat[][LP~1033-31]{\includegraphics[width=11.5cm, angle=0]{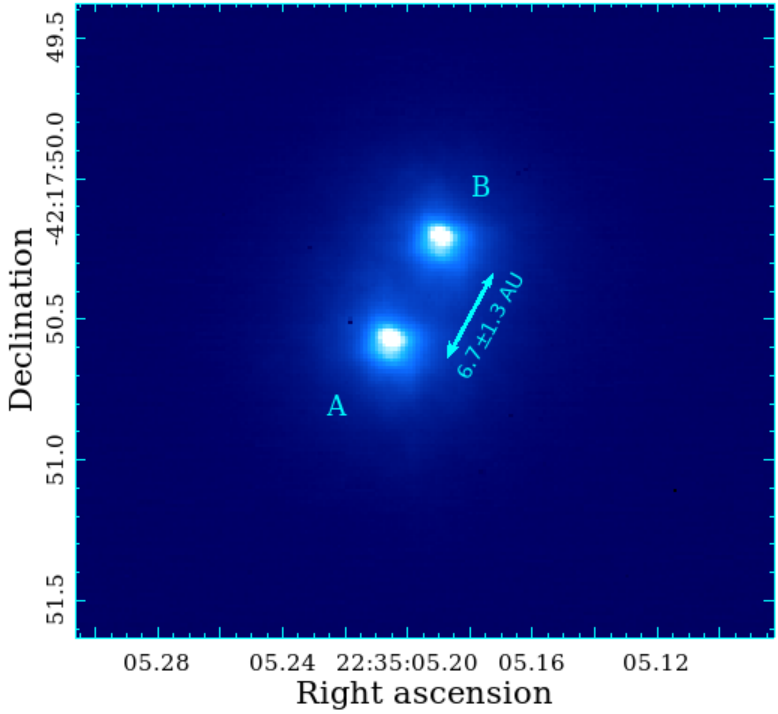}\label{fig:1a}}
  
    \subfloat[][LP~877-72]{\includegraphics[width=11.5cm, angle=0]{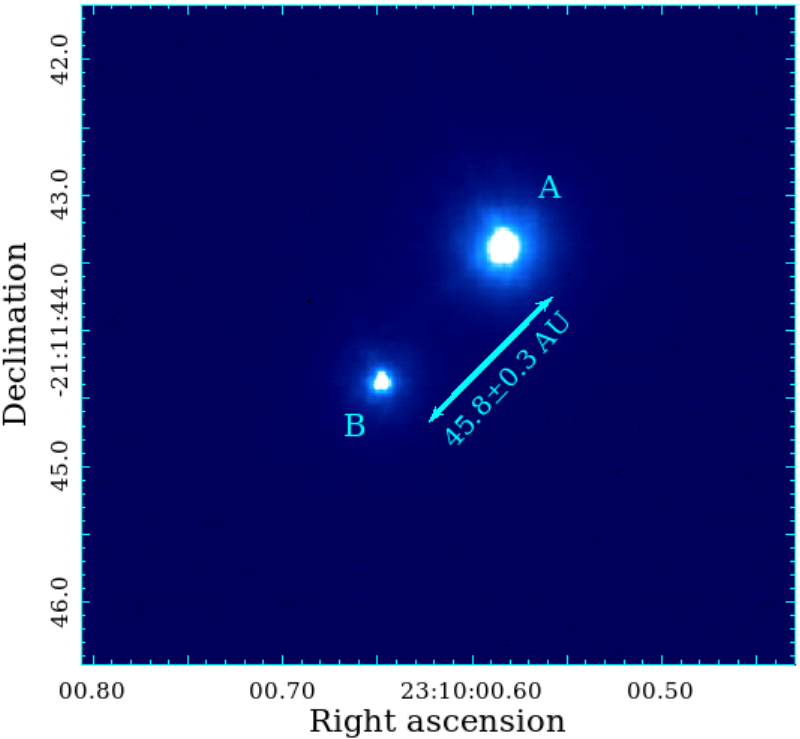}\label{fig:1b}}
	\caption{The $J$-band images obtained from VLT/NaCo instruments: (a) LP 1033-31 and (b) LP 877-72. The standard conversion has been used, i.e., north is up, and east is left. A and B denote the primary and secondary components of the VLM binaries.}
	\label{fig:fig_1}
\end{figure*}

\section{Data Analysis}
In this research, we used the raw image mode data that were provided in ESO archive\footnote{\href{https://archive.eso.org/eso/eso_archive_main.html}{https://archive.eso.org/eso/eso\_archive\_main.html}}. The data reduction was performed using the \textit{Image Reduction and Analysis Facility} \citep[\iraf\footnote{\href{http://iraf.net}{http://iraf.net}}; ][]{TodyD-86-SPIE, TodyD-93-ASPC} software. In order to perform various operations, including source coordinate detection, position angle, and separation measurements, we have used python packages, including {\sc photutils}\footnote{\href{https://github.com/astropy/photutils/tree/1.6.0}{https://github.com/astropy/photutils/tree/1.6.0}} \citep{BradleyL-22-zndo-1}, {\sc pyastronomy}\footnote{\href{https://github.com/sczesla/PyAstronomy/}{https://github.com/sczesla/PyAstronomy/}} \citep[][]{CzeslaS-19-ascl-2}, and {\sc astropy}\footnote{\href{https://github.com/astropy/astropy}{https://github.com/astropy/astropy}} \citep[][]{Astropy-Collaboration.-22-ApJ}. In order to perform the near-infrared (NIR) photometry in $JHK_s$ photometric bands, we followed the standard procedures. Please see \cite{KarmakarS-20-MNRAS-2} for more details about the reduction methods. 

\begin{table*}
\center
\small
\caption{\label{tab:parameters} System parameters for \lp1\ and \lp8}
\setlength{\tabcolsep}{6mm}
\begin{tabular}{l|cc|cc}
\hline
\hline
\noalign{\smallskip}
\multirow{2}{*}{Parameters} & \multicolumn{2}{c|}{\lp1}& \multicolumn{2}{c}{\lp8} \\
& A & B & A & B\\

\noalign{\smallskip}
\hline
\hline
\noalign{\smallskip}
 SpT$^{\dagger}$ & M4.5$\pm$0.5 & M4.5$\pm$0.5 & M1.0$\pm$0.3 & M3.8$\pm$0.3\\[0.4mm]
Mass (\msun)$^{\dagger \dagger}$ & 0.20$\pm$0.04 & 0.19$\pm$0.04 & 0.520$\pm$0.006 & 0.260$\pm$0.005 \\[0.4mm]
Radius (\rsun) & 0.225$\pm$0.030 & 0.217$\pm$0.029 & 0.492$\pm$0.011 & 0.270$\pm$0.006 \\[0.4mm]
$\teff$ (K) & 3245$\pm$107 & 3211$\pm$120 & 3750$\pm$15 & 3348$\pm$10 \\[0.4mm]
$\logg$ $^{\ddagger}$& 5.061$\pm$0.038 & 5.068$\pm$0.039 & 4.768$\pm$0.005 & 4.984$\pm$0.004 \\[0.4mm]
 Luminosity (\lsun) & 0.0050$\pm$0.0015& 0.0045$\pm$0.0014 & 0.0432$\pm$0.0021 & 0.0082$\pm$0.0004 \\[0.4mm]

\noalign{\smallskip}
\hline
\hline
\end{tabular}
\\[1mm]
\scriptsize
$^{\dagger}$ -- For the individual systems, the spectral types have been derived from M$_{J}$ using the method of  \cite{Scholz-05-A+A-9}; $^{\dagger \dagger}$ -- Mass is derived from the interpolation of \cite{BaraffeI-15-A+A} isochrones of 5 Gyr;  $^{\ddagger}$ -- $\logg$ is the logarithmic value (base 10) of surface gravity~($g$), where $g$ is in cm~s$^{-2}$ unit.\\
\normalsize
\end{table*}

\section{Brightness and Binary Separation}
Fig.~\ref{fig:fig_1} shows the close-up view of the representative $J$-band images for the objects \lp1\ and \lp8. In the figure, the usual conventions were followed, i.e., north is up, and east is left. Both objects are clearly resolved as binary systems. In our analysis, the brighter component is considered primary. Fig.~\ref{fig:1a} indicates the \lp1\ system containing two components with the measured $J$-band magnitudes of 9.82$\pm$0.04 and 9.90$\pm$0.04. The components have almost equal brightness with a slightly higher flux of the South-East component (\lp1~A) than the North-West component (\lp1~B). For the \lp8\ system, the $J$-band magnitudes are derived as 9.08$\pm$0.03 for the primary (\lp1~A), and 10.77$\pm$0.04 (\lp8~B) for the secondary component. The binary separations for the \lp1\ and \lp8\ systems have been estimated to be 402$\pm$1 and 1335$\pm$2 mas, respectively. Following \cite{Finch-14-AJ-8} and \cite{Bailer-Jones-18-AJ-6}, in this work, we have adopted the distances for \lp1\ and \lp8\ to be 16.57$\pm$3.27 and 33.77$\pm$0.17 pc, respectively. Therefore, the binary separation between the primary and secondary components of \lp1\ is computed as 6.7$\pm$1.3~AU. Similarly, the binary separation between \lp8~A and B is derived as 45.8$\pm$0.3~AU.

\section{Absolute magnitude and Spectral Type}
The absolute $J$-band magnitude for the primary and secondary components of \lp1\ are derived to be 8.7$\pm$0.4 and 8.8$\pm$0.4, respectively. On the other hand, for the primary and secondary components of \lp8, the absolute $J$-band magnitudes are derived as 6.44$\pm$0.03 and 8.13$\pm$0.04, respectively. From the Hertzsprung-Russell Diagram, it is possible to determine the spectral type for each of the components of both systems. \cite{Scholz-05-A+A-9} investigated the relationship between absolute $J$ magnitude and the spectral type for K0--L8 stars. From the computed absolute magnitude, we derived the spectral types of the systems \lp1\ and \lp8\ are M4.5+M4.5 and M1+M4, respectively. For comparison, the latest estimated spectral types of the `unresolved' combined systems of \lp1\ and \lp8, as derived by \cite{Rajpurohit-13-A+A-2}, are given as M4 and M4, respectively. Currently, the spectra of the individual components have not been observed in any of the systems. However, the combined spectra of \lp1\ and \lp8\ have already been observed by \cite{Reyle-06-MNRAS-3}. We also performed a combined spectral fitting using the latest available empirical template library of stellar spectra from \cite{Kesseli.-17-ApJS-2}. We fitted the combined spectra at the flux ratios derived in our analysis. The spectral fitting of the individual components with the derived spectral types is found to be better than the best-fitted template spectra of the unresolved systems. Please also see \cite{KarmakarS-20-MNRAS-2} for a detailed description. 

\section{Stellar parameters}
Stellar magnetic activities are found to be dependent on spectral type as well as stellar age. Investigating 38,000 low-mass stars from the Sloan Digital Sky Survey (SDSS) Data Release 5 (DR5), \cite{West-08-AJ-7} have derived `activity lifetimes' or ages for each spectral types of M0--M7 dwarfs. As an M4.5+M4.5 binary, the individual components of \lp1\ should have an age of 3.5--7.5 Gyr (since the age of M4 and M5 stars are 4.5$^{+0.5}_{-1.0}$ and 7.0$\pm$0.5 Gyr). On the other hand, the spectral types of \lp8~A and B components are M1 and M4, which for a `single' un-related systems, according to age-activity relations, may have ages $<$0.8 and 4.5$^{+0.5}_{-1.0}$ Gyr, respectively. However, for a physically bound system, it is more likely that both systems are coeval. Further, from the analysis of evolutionary tracks, it has been shown that there are only little significant differences between evolutionary tracks of 0.6--10.0~Gyr \citep[][]{BaraffeI-15-A+A}. Therefore, we conservatively assume the age of $\sim$5 Gyr. We have estimated the masses of both the components of \lp1\ and \lp8\ using the mass-absolute magnitude relation as derived by \cite{BenedictG-16-AJ}.  \lp1~AB are found to have the nearly equal mass of 0.20$\pm$0.04 and 0.19$\pm$0.04~\msun, whereas, for \lp8~AB, the derived masses are found to be 0.520$\pm$0.006 and 0.260$\pm$0.005~\msun, respectively. Using the mass-radius relations of \cite{Boyajian-12-ApJ-3}, the radii of \lp1~A and B are derived to be 0.225 $\pm$ 0.030 and 0.217 $\pm$ 0.029 \rsun. On the other hand, for LP 877-72 AB, the radii are estimated to be 0.492 $\pm$ 0.011 and 0.270 $\pm$ 0.006~\rsun\ for primary and secondary, respectively. 
Using the measured projected separations and the masses, we estimated the orbital periods for each of the systems from \textit{Kepler}'s law. For \lp1\ and \lp8, the orbital periods are derived to be 28$\pm$3 and 349$\pm$3 yrs. 
For \lp1, the $\teff$ and $\logg$ are found to be similar, whereas the primary of \lp8\ is derived to be $\sim$400 K hotter and the surface gravity is lower by a factor of $\sim$1.6 than the secondary component. Please see  Table~\ref{tab:parameters} for the estimated parameters of detected binary components.

\section{Discussion}
Using high-resolution adaptive optics NIR $JHK_s$ band images from VLT/NaCo instruments, in this work, we investigated two VLM systems, \textit{viz.}  \lp1\ and \lp8. We found that \lp1\ consists of two components with similar brightness, mass, radius, surface temperature, log g, and luminosity. However, in the case of \lp8, these parameters vary between the primary and secondary over a wide range. A detailed discussion is beyond the scope of this article. Please find the paper \cite{KarmakarS-20-MNRAS-2} for a robust discussion.

 In the remaining part of this section, we will briefly discuss the possibility of hosting exoplanets on the identified VLM components. The radial velocity studies indicate that M-dwarfs host at least 2.39$^{+4.58}_{-1.36}$ planets per star orbiting them \citep{TuomiM-19-arXiv}. Using \textit{Kepler} data, \cite{Hardegree-Ullman-19-AJ-2} estimated the planet occurrence rate for spectral type ranges M3 V -- M3.5 V, M4 V -- M4.5 V, and M5 V -- M5.5 V to be 0.86$^{+1.32}_{-0.68}$, 1.36$^{+2.30}_{-1.02}$, and 3.07$^{+5.49}_{-2.49}$ planets per star. In the case of the binary systems, close binary companions appear to suppress planet formation and hence decrease planet occurrence rates for these systems \citep[see][]{WangJ-14-ApJ-24, KrausA-16-AJ-2}. Considering the suppression due to the binarity, we derive that the probability of occurring planet for each of the primary and secondary components of \lp1~AB would be the same and with the range of 4--51\%. In the case of \lp8, the primary and secondary components of the planet occurrence rate would be $\lsimeq$ 49\% and 19--208\%. In other words, while searching for exoplanets around each of the components of the VLM binaries, we should expect to find up to `two' exoplanets around \lp8~B, in contrast with the \lp8~A, \lp1~A, and \lp1~B, all three of the components have the maximum probability of hosting exoplanet are only $\sim$50\%. 

\section*{Acknowledgments}
{
This research is based on the observations obtained with the NaCo instrument on the VLT@ESO telescope at Paranal Observatory under program ID 091.C-0501(B). This proceeding article is a derivative of \cite{KarmakarS-20-MNRAS-2}, which includes the contributions of France Allard. With our deepest sorrow, France Allard was deceased on October 2020. We dedicate this paper to France Allard.
}

\bibliographystyle{cs21proc}
\bibliography{SK_collections.bib}

\end{document}